\documentclass{elsart}

\usepackage{amsfonts}
\usepackage{graphicx}
\usepackage{bm}

\def\slashchar#1{\setbox0=\hbox{$#1$}
   \dimen0=\wd0 \setbox1=\hbox{/} \dimen1=\wd1
   \ifdim\dimen0>\dimen1 \rlap{\hbox to \dimen0{\hfil/\hfil}} #1
   \else  \rlap{\hbox to \dimen1{\hfil$#1$\hfil}} / \fi}

\begin{document}

\begin{frontmatter}

\title{Approximating chiral quark models with linear 
$\sigma$-models\thanksref{grants}}
\thanks[grants]{Research supported in part by the
Scientific and Technological Cooperation Joint Project between
Poland and Slovenia, financed by the
Ministry of Science of Slovenia and the Polish State Committee for
Scientific Research, and by the Polish State Committee for
Scientific Research, grant number 2 P03 09419}
\thanks[emails]{wojciech.broniowski@ifj.edu.pl, 
bojan.golli@ijs.si}
\author[INP]{Wojciech Broniowski} and
\author[LJ]{Bojan Golli}
\address[INP]{The H. Niewodnicza\'nski Institute of Nuclear
Physics, ul. Radzikowskiego 152, PL-31342 Krak\'ow, Poland}
\address[LJ]
{Faculty of Education, University of Ljubljana and J.~Stefan Institute,
Jamova 39, P.O.~Box 3000, 1001 Ljubljana, Slovenia}

\date{2 October 2002}

\begin{abstract}
We study the approximation of chiral quark models with 
simpler models, obtained via gradient expansion. 
The resulting Lagrangian of the type of the
linear $\sigma$-model contains, at the lowest level of the 
gradient-expanded meson action, an additional term of the form
$\frac{1}{2} A (\sigma \partial_\mu \sigma + \pi \partial_\mu \pi )^2$. 
We investigate the dynamical consequences of this term and its relevance to
the phenomenology of the soliton models of the nucleon. 
It is found that the inclusion of the
new term allows for a more efficient approximation 
of the underlying quark theory, especially 
in those cases where dynamics allows for a large deviation of
the chiral fields from the chiral circle, such as in quark models
with non-local regulators. This is of practical importance, since the $\sigma$-models 
with valence quarks only are technically much easier to treat and simpler to solve 
than the quark models with the full-fledged Dirac sea.
\end{abstract}

\begin{keyword} Effective chiral quark models,
$\sigma$-model, Nambu--Jona-Lasinio model, chiral hedgehog solitons
\end{keyword}

\end{frontmatter}

\noindent PACS: 12.39.-x, 12.39.Fe, 12.40.Yx

\section{Introduction}\label{sec:intro}

Various $\sigma$-models, \emph{i.e.} effective models incorporating the
spontaneously broken chiral symmetry, play an important role in modeling the
low-energy hadronic physics. The original Gell-Mann-L\'{e}vy linear 
$\sigma$-model \cite{GML}, after replacing nucleons by quarks 
\cite{BirBan8485,KRS84,KR84,EisKal,BB8586,McGovern}, 
has been used to describe the
binding mechanism of quarks inside baryons and to produce quite successful
phenomenologies of baryons. Actually, these models, with Lagrangians
containing up to two derivatives of fields, may be viewed as formal
approximations to models with purely quark degrees of freedom, such as the
Nambu--Jona-Lasinio (NJL) model \cite{NJL} with quarks (for review and further 
references to these models in the context of mesons see, {\em e.g.}, 
\cite{Vogl2,Klevansky,Hatsuda94}, and in the contexts of chiral solitons
{\em e.g.},
\cite{NJL:rev,Alkofer:rev,mitja:rev,ripka:book}). Indeed, gradient
expansion techniques \cite{Eguchi:boson,Aitchison,Nepomechi,Zuk} 
produce mesonic terms of
the form used in various $\sigma $-models. Admittedly, questions of
convergence of the gradient-expansion series are difficult, and it is not
clear how many terms should be present in the mesonic theory such that it
approximates satisfactorily the underlying quark theory. On the other hand,
the $\sigma$-like models are much simpler to solve, since they do not carry
the complications of the Dirac sea, which for models of baryons is
technically difficult to treat, and a massive computational effort 
is involved. 
Therefore, it is desirable to
understand how to construct much simpler and more intuitive 
effective $\sigma$-models which carry 
valence quarks only, and then to investigate
what role is played by various mesonic terms arising from 
the gradient expansion.

In this paper we analyze a particular extension of the linear $\sigma $%
-model coupled to valence quarks, which contains an additional term with two
gradients of the chiral fields. This term, arising naturally in the process
of gradient-expanding NJL-like models, has not been, to our
knowledge, considered before. It
has dynamical consequences for the model and, in addition, allows for a
more efficient approximation of the underlying quark model. We discuss the
nature of the term and its consequences for the Euler-Lagrange equations in
Sec. \ref{sec:A}. 
In Sec. \ref{sec:motiv} we show 
how the term arises when the NJL model is gradient-expanded 
and explain how the parameters of 
the resulting $\sigma$-model can be derived.
%%, and estimate its strength. 
%%For that purpose we use the
We estimate the strength of the new term in 
NJL model with the sharp four-momentum cut-off, and find that it is
large. We then proceed to analyzing its dynamical effects for chiral
solitons. In Sec. \ref{sec:sol} we consider the straightforward 
extension of the 
$\sigma $-model, and find that the new term lowers the energy 
of the soliton,
thus contributing to binding and its energetic stability. 
In Sec. \ref{sec:nonloc} we apply our analysis to the chiral quark model
with non-local regulator 
\cite{mitja:rev,ripka:book,Diak86,DPPob88,PlantB,BowlerB,solnonl,%%
solilong,Carter,Michal1,Dorokhov:nl},
where 
we show that the inclusion of the new term allows for a more 
efficient approximation of the 
underlying quark theory. 

\section{The $A$-term}\label{sec:A}

By means of general symmetry arguments, 
in the construction of the effective Lagrangian one has at hand the
invariants of the given symmetry group, in the case considered here 
the chiral $SU(2)_{L}\times
SU(2)_{R}$. The chiral fields, \emph{i.e.} the scalar-isoscalar $\sigma$-field
and the three pion fields, $\pi ^{a}$, belong to the $(1/2,$ $1/2)$ 
representation of this
group. The typical chiral invariants one can form out of the chiral fields,
arranged in the increasing number of spatial derivatives, are 
\begin{eqnarray}
&& \sigma ^{2}+\pi ^{2},\quad  \nonumber \\
&&\sigma \partial ^{\mu }\sigma +\pi ^{a}\partial ^{\mu }\pi ^{a},  
\nonumber \\
&&(\partial ^{\mu }\sigma )(\partial ^{\nu}\sigma )+(\partial ^{\mu }\pi
^{a})(\partial ^{\nu }\pi ^{a}),  \nonumber \\
&&(\partial ^{\mu \alpha }\sigma )(\partial ^{\nu}\sigma )+(\partial ^{\mu
\alpha }\pi ^{a})(\partial ^{\nu }\pi ^{a}), \\
&&{\rm etc.}  \nonumber
\end{eqnarray}
Similarly, the coupling to Dirac fields, $\psi$, allows to construct
invariants of the form $\bar{\psi}(\sigma +i\gamma _{5}\tau ^{a}\pi
^{a})\psi $, $\bar{\psi}\gamma _{\mu }(\sigma \partial ^{\mu }\sigma +\pi
^{a}\partial ^{\mu }\pi ^{a})\psi $, {\em etc.} The above invariants are
next combined in such a way as to form Lorentz invariants. In the mesonic
sector, up to terms involving two gradients, we have the general form for
the mesonic Lagrangian, 
\begin{equation}
L_{\mathrm{mes}}=-V+\frac{1}{2}Z[(\partial _{\mu }\sigma )^{2}+(\partial
_{\mu }\pi ^{a})^{2}]+\frac{1}{2}A[\sigma \partial ^{\mu }\sigma +\pi
^{a}\partial ^{\mu }\pi ^{a}]^{2},  \label{Lag}
\end{equation}
with $V$, $Z$, and $A$ denoting, at the moment, 
arbitrary functions of the invariant $\sigma ^{2}+\pi ^{2}$. 
The last term in Eq. (\ref{Lag}), called from now on the $A$-term, is 
new compared to previous works. In the original Gell-Mann--L\'evy
model this term is excluded by the requirement of renormalizability.
Since we are going to use (\ref{Lag}) as an effective model,
approximating the underlying quark theory, the model need not and should
not be renormalizable, hence the $A$-term may be present.
Moreover, as we will show in
Sec. \ref{sec:motiv}, the appearance of the $A$-term naturally 
follows from the gradient
expansion of the quark models with the Dirac sea, such as the NJL model
and its modifications.

The linearity of the model is an essential feature in our study. The
non-linear constraint imposed on the chiral fields, $\sigma ^{2}+\pi ^{2}=const.$,
immediately kills the $A$-term, since $\sigma \partial ^{\mu }\sigma +\pi
^{a}\partial ^{\mu }\pi ^{a}=\frac{1}{2}$ $\partial ^{\mu }(\sigma ^{2}+\pi ^{2})=0$. 
As a matter of fact, the underlying quark models lead to linear chiral
models, and the non-linearity in most approaches \cite{NJL:rev,Alkofer:rev,mitja:rev,ripka:book} 
is superimposed externally.
\footnote{In fact, in these works the non-linear constraint is needed to provide 
stability of solitons \cite{NJL:rev,Alkofer:rev}. On the other hand, in non-local 
models of Refs. \cite{solnonl,solilong} stability of solitons is achieved
without the ``external'' non-linear constraint.}

In the following we shall use the coupling to quarks of the form as in the
Gell-Mann-Levy model \cite{GML}, \emph{i.e.} $-g\bar{\psi}(\sigma +i\gamma _{5}\tau
^{a}\pi ^{a})\psi $, with $\psi $ containing the valence quark orbit only. The Dirac sea has
been (approximately) integrated out and entered into the mesonic part of the Lagrangian. In
addition, for simplicity of the calculations, we assume that the functionals 
$Z$ and $A$ are constant, $i.e.$ do not depend on the value of the chiral fields: 
\begin{equation}
Z(\sigma ^{2}+\pi ^{2}) =Z_{0},  \;\;\;\;\; A(\sigma ^{2}+\pi ^{2}) =A_{0}.
\end{equation}
This assumptions is justified as long as the field do not depart too far away from
the chiral circle.

The mean-field Euler-Lagrange equations resulting from Eq. (\ref{Lag}) have the form 
\begin{eqnarray}
Z_{0}\Box \sigma +A_{0}\sigma \left[ (\partial _{\mu }\sigma )^{2}+(\partial
_{\mu }\pi ^{a})^{2}+\sigma \Box \sigma +\pi ^{a}\Box \pi ^{a}\right] +\frac{%
\partial V}{\partial \sigma }+j_{\sigma } &=&0,  \nonumber \\
Z_{0}\Box \pi ^{a}+A_{0}\pi ^{a}\left[ (\partial _{\mu }\sigma
)^{2}+(\partial _{\mu }\pi ^{b})^{2}+\sigma \Box \sigma +\pi ^{b}\Box \pi
^{b}\right] +\frac{\partial V}{\partial \pi ^{a}}+j_{\pi}^{a} &=&0,
\label{EL0}
\end{eqnarray}
where $j_{\sigma }$ and $j_{\pi}^{a}$ are the valence quark sources, 
\begin{equation}
j_{\sigma} =g\sum_{i\in val}\bar{q}_{i}q_{i},  \;\;\;\;\; 
j_{\pi}^{a} =g\sum_{i\in val}\bar{q}_{i}i\gamma _{5}\mathbf{\tau }^{a}q_{i},  \label{qsou}
\end{equation}
and $q_{i}$ are the valence orbitals.\footnote{The effects of the Dirac sea are included in the
mesonic degrees of freedom.} As can be seen from Eqs. (\ref{EL0}), the $A$%
-term mixes the propagation of the chiral fields. Equations (\ref{EL0}) can
be diagonalized and brought to the simple matrix form 
\begin{eqnarray}
&&\left( 
\begin{array}{c}
\Box \sigma \\ 
\Box \pi ^{a}
\end{array}
\right) =-\frac{1}{Z_{0}\left( Z_{0}+(\sigma ^{2}+\pi ^{2})A_{0}\right) }%
\times   \label{ELmat}  \\ \nonumber
&&\ \ \ \left( 
\begin{array}{cc}
Z_{0}+A_{0}\pi ^{2} & -A_{0}\sigma \pi ^{a} \\ 
-A_{0}\sigma \pi ^{a} & Z_{0}+A_{0}\sigma ^{2}
\end{array}
\right) \left( 
\begin{array}{c}
A_{0}\sigma ((\partial _{\mu }\sigma )^{2}+(\partial _{\mu }\pi
^{b})^{2})+\partial V/\partial \sigma +j_{\sigma } \\ 
A_{0}\pi ^{a}((\partial _{\mu }\sigma )^{2}+(\partial _{\mu }\pi
^{b})^{2})+\partial V/\partial \pi ^{a}+j_{\pi}^{a}
\end{array}
\right) .  
\end{eqnarray}
If $A_{0}=0$, the usual Gell-Mann-L\'evy $\sigma$-model equations 
of motion are recovered.

Equations (\ref{EL0}) are simple to solve in the vacuum sector. 
In this case the $A$ and $Z$ terms do not contribute and one has
the familiar result:
\begin{equation}
\langle \sigma \rangle =F,  \label{vac} \;\;\;\;\; \langle \pi ^{a}\rangle =0, 
\end{equation}
where $F=93$~MeV denotes the pion decay constant. As usual, we introduce the
shifted $\sigma $-field, $\sigma ^{\prime }=\sigma -F$, and find for the
small-amplitude fluctuations of the chiral fields around their vacuum values
(\ref{vac}) the following equations of motion
\begin{eqnarray}
(Z_{0}+A_{0}F^{2})\Box \sigma ^{\prime }+\left. \frac{\partial ^{2}V}{%
\partial \sigma ^{2}}\right| _{\sigma ^{\prime }=\pi =0}\sigma ^{\prime }
&=&0,  \nonumber \\
Z_{0}\Box \pi ^{a}+\left. \frac{\partial ^{2}V}{\partial \pi ^{a}\partial
\pi ^{b}}\right| _{\sigma ^{\prime }=\pi =0}\pi ^{b} &=&0. 
\end{eqnarray}
We can thus canonically normalize the pion field by choosing $Z_{0}=1$.
Then, for non-zero $A_{0}$, the normalization of the $\sigma $ field is not
canonical, and carries the factor $1+A_{0}F^{2}$, which means that the mass
squared of the physical $\sigma ^{\prime }$ field is given by the expression 
\begin{equation}
m_{\sigma }^{2}=(1+A_{0}F^{2})^{-1}\left. \frac{\partial ^{2}V}{\partial
\sigma ^{2}}\right| _{\sigma ^{\prime }=\pi =0}  \label{ms}
\end{equation}

One can easily check that the $A$-term \emph{does not contribute} explicitly
to the vector, $V_{\mu }^{a}$, and axial-vector, $A_{\mu }^{a}$, Noether
currents. The meson contributions remain to be the usual $\sigma $-model
expressions, 
\begin{eqnarray}
V_{\mu }^{a} &=&Z\epsilon ^{abc}(\partial _{\mu }\pi ^{b})\pi ^{c}, 
\nonumber \\
A_{\mu }^{a} &=&Z\left( \sigma \partial _{\mu }\pi ^{a}-\pi ^{a}\partial
_{\mu }\sigma \right) .  \label{VA}
\end{eqnarray}
The above expressions hold for the case of general $Z$ and $A$. 
Therefore the expressions for the magnetic moments or $g_A$ used in 
soliton calculations \cite{BC86} are not modified.
However, changes in these quantities are induced by the $A$-term through
the change of the dynamics in Eqs. (\ref{ELmat}). 

\section{Motivation from quark models}\label{sec:motiv}

We are now going to argue that the presence of the $A$-term in effective
chiral Lagrangians follows naturally if one considers approximations to
chiral quark models, such as the NJL model. For simplicity of notation we work
in the strict chiral limit of the vanishing current quark mass. The action of
the bosonized NJL model at the one-quark-loop level can be written as
(see, {\em e.g.}, \cite{NJL:rev,Alkofer:rev,ripka:book})
\begin{eqnarray}
I(\sigma ,\pi ^{a}) &=&-Tr\log \left( \partial _{\tau }+h\right) -\frac{g^{2}%
}{2G^{2}}\int d^{4}x(\sigma ^{2}+\pi ^{2}),  \label{Action} \\
h &=&-i\vec{\alpha}\cdot \vec{\nabla}+\beta g(\sigma +i\gamma _{5}\tau
^{a}\pi ^{a}).  \label{ham0}
\end{eqnarray}
The trace is over color, flavor, Dirac, and functional space, $\tau $
denotes the Euclidean time.
We have introduced an additional parameter, $g$, 
the quark-meson coupling constant, which allows us to fix the vacuum 
value of the $\sigma$-field by imposing the condition (\ref{vac}).
The constant $G$ is the four-quark coupling constant of the original NJL model
\cite{NJL}. The constituent quark mass is related to the pion decay constant, 
\begin{equation}
M=gF.  \label{GT}
\end{equation}
For many practical applications (study
of large-size solitons, checking numerical computations) it is useful to
have an approximation to this model, which is much simpler to deal
with. In practice, this amounts to the gradient expansion 
\cite{Eguchi:boson,Aitchison,Nepomechi,Zuk} 
of action (\ref{Action}), with the hope that for sufficiently 
large-size solitons
the expansion should work fine.
Also, in many models the fields lie close to the
chiral circle, therefore another good expansion parameter is the combination 
\begin{equation}
\delta =\sigma ^{2}+\pi ^{2}-F^{2},  \label{x}
\end{equation}
which measures the departure of the fields from the chiral circle. We can
thus expand 
\begin{equation}
V(\delta )=\sum_{n=0}^{\infty }V_{n}\delta ^{n},\qquad Z(\delta
)=\sum_{n=0}^{\infty }Z_{n}\delta ^{n},\qquad A(\delta )=\sum_{n=0}^{\infty
}A_{n}\delta ^{n}. \label{expan}
\end{equation}
The coefficients $V_{n}$, $Z_{n}$ and $A_{n}$ can be found by comparing the
Green's functions obtained from (\ref{Action}) to those obtained at 
the effective
level from (\ref{Lag}).

Our technique may be viewed as a double expansion: in the number of 
gradients, up to two, and
in the parameter $\delta$. In the functions $Z$ and $A$ we thus keep 
the terms up to zeroth
order in $\delta$, while in the potential $V$ we shall keep term up to 
second order in $\delta$. 

To begin, we recall the gradient expansion of $V$, which is
easy, since for that purpose one can treat the fields as space-time
independent. We also have the grand-reversal symmetry \cite{CB86}, 
which leads to 
$\mathrm{Tr}\log \left( \partial _{\tau }+h\right) =$ 
$\mathrm{Tr}\log \left(
-\partial _{\tau }+h\right) =\frac{1}{2}\mathrm{Tr}\log \left( -\partial
_{\tau }^{2}+h^{2}\right) $. Therefore 
\begin{eqnarray}
V(\sigma ^{2}+\pi ^{2})&=&-2N_{c}N_{f}\int \frac{d_{4}k}{(2\pi )^{4}}\log
\left( k^{2}+g^{2}\left( \sigma ^{2}+\pi ^{2}\right) \right) +\frac{g^{2}}{%
2G^{2}}(\sigma ^{2}+\pi ^{2}) \nonumber \\ &+&\mathrm{const.,} \label{V}
\end{eqnarray}
where $k$ is the Euclidean momentum flowing along the quark loop, $N_{c}=3$
is the number of colors, and $N_{f}=2$ is the number of flavors. The
constant can be chosen such that $V=0$ in the vacuum (\ref{vac}). In
addition, the stationary-point condition, $\left. \partial V/\partial \sigma
\right| _{\rm vac}=0$, has the explicit form 
\begin{equation}
\frac{1}{2G^{2}}=2N_{c}N_{f}\int \frac{d_{4}k}{(2\pi )^{4}}\frac{1}{%
k^{2}+M^{2}}. \label{gap}
\end{equation}
Plugging (\ref{gap}) into (\ref{V}) results in the following formula for $V$: 
\begin{equation}
V(\delta )=-2N_{c}N_{f}\int \frac{d_{4}k}{(2\pi )^{4}}\left[ \log \left( 1+%
\frac{g^{2}\delta }{k^{2}+M^{2}}\right) -\frac{g^{2}\delta }{k^{2}+M^{2}}%
\right]. \label{Vexp}
\end{equation}
The coefficients of the expansion (\ref{Vexp}) are: 
\begin{eqnarray}
V_{0} &=&V_{1}=0,  \nonumber \\
V_{n} &=&\frac{(-)^{n}}{n}2N_{c}N_{f}\int \frac{d_{4}k}{(2\pi )^{4}}\left( 
\frac{1}{k^{2}+M^{2}}\right) ^{n},\qquad n>1.  \nonumber
\end{eqnarray}
We should bear in mind that the expansion for the $\log $ is rather slowly
convergent, such that many terms in Eq. (\ref{expan}) may be needed 
to reproduce
well the full result (\ref{V}).
Retaining the $n=2$ term only corresponds to the use of the 
usual Mexican hat potential.

Next, we calculate the coefficients $Z_{0}$ and $A_{0}$. For that
purpose one can consider mesonic two-point Green functions. 
In the vacuum the inverse $%
\sigma $ and pion propagators obtained from Eq. (\ref{Lag}) 
are (the momentum $q$ is Euclidean): 
\begin{eqnarray}
K_{\pi }^{-1}(q)\delta ^{ab} &\equiv &\frac{\delta ^{2}L}{\delta \pi
^{a}(q)\delta \pi ^{b}(-q)}=Z_{0}q^{2}\delta ^{ab},  \label{propL} \\
K_{\sigma }^{-1}(q) &\equiv &\frac{\delta ^{2}L}{\delta \sigma (q)\delta
\sigma (-q)}=\left. \frac{\partial ^{2}V(\sigma ^{2}+\pi ^{2})}{\partial
\sigma ^{2}}\right| _{\mathrm{vac}}+\left( Z_{0}+F^{2}A_{0}\right) q^{2}. 
\nonumber
\end{eqnarray}
Meanwhile, in the NJL model one obtains, with help of 
Eqs. (\ref{Action},\ref{ham0}),  
the following formulas for the inverse meson propagators: 
\begin{eqnarray}
K_{\pi }^{-1} &=&2g^{2}N_{c}N_{f}\ f(M,q^{2})q^{2}=2g^{2}N_{c}N_{f}\
f(M,0)q^{2}+O(q^{4}),  \label{propNJL} \\
K_{\sigma }^{-1} &=&2g^{2}N_{c}N_{f} f(M,q^{2})(q^{2}+4M^{2})  
 = 8g^{2}N_{c}N_{f}M^{2}\ f(M,0)\nonumber \\ &+&2g^{2}N_{c}N_{f}\ \left(
f(M,0)+4M^{2}\left. \ df(M,q^{2})/dq^{2}\right| _{q=0}\right) q^{2}+O(q^{4}),
\nonumber
\end{eqnarray}
where 
\begin{equation}
f(M,q^{2})=\int_{\Lambda }\frac{d_{4}k}{(2\pi )^{4}}\frac{1}{\left( \left(
k+q/2\right) ^{2}+M^{2}\right) \left( \left( k-q/2\right) ^{2}+M^{2}\right) }%
.  \label{fM}
\end{equation}
The index $\Lambda $ in the integral denotes an appropriate 
regulator \cite{NJL},
necessary for the model to become finite. By comparing (\ref{propL}) 
and (\ref{propNJL}) we immediately arrive at the relations 
\begin{eqnarray}
Z_{0} &=&2g^{2}N_{c}N_{f}f(M,0), \\
A_{0} &=&8g^{2}N_{c}N_{f}(M/F)^{2}\left. \ df(M,q^{2})/dq^{2}\right| _{q=0}\
.  \nonumber
\end{eqnarray}
Through the use of the relation $F=M\sqrt{2N_{c}N_{f}f(M,0)}$ \cite{NJL:rev} 
and Eq. (\ref{GT}) we obtain the formulas 
\begin{eqnarray}
Z_{0}&=&1 \quad\mbox{or}\quad 
 g^{-2} = 2N_{c}N_{f}f(M,0),
\\
A_{0}&=&4g^{2}\frac{\left. \ df(M,q^{2})/dq^{2}\right| _{q=0}}
{f(M,0)} \nonumber .
\end{eqnarray}

In order to estimate the size of the $A$-term, in Fig. 1 we plot the
quantity $1+F^{2}A_{0}$, which plays the role of the wave function 
renormalization for the $%
\sigma$-field. In the study of this section we have used the sharp 
four-momentum cut-off
on the quark loop in Eq. (\ref{fM}). The calculation is performed in the 
chiral limit, and with the cut-off $\Lambda$ chosen in such a way as to 
reproduce the physical value of $F=93$~MeV. 
We note that the effect of the $A$-term
is very strong, with the combination $1+F^{2}A_{0}$ ranging from 0.6 at 
$M=300$~MeV to as little as 0.1 at $M=600$~MeV, 
to be compared to unity of the case 
without the $A$-term.

It is easy to derive full expressions for $Z$ and $A$. With the notation 
of Eq. (\ref{x}) we find
\begin{eqnarray}
Z(\delta ) &=&2g^{2}N_{c}N_{f}\int \frac{d_{4}k}{(2\pi )^{4}}\frac{1}{\left(
k^{2}+\left( g^{2}\delta +M^{2}\right) \right) ^{2}}, \\
A(\delta ) &=&-2g^{4}N_{c}N_{f}\int \frac{d_{4}k}{(2\pi )^{4}}\frac{%
k^{2}+2\left( g^{2}\delta +M^{2}\right) }{\left( k^{2}+\left( g^{2}\delta
+M^{2}\right) \right) ^{4}}.
\end{eqnarray}
These formulas can be straightforwardly expanded in powers of $\delta $ to
produce all coefficients in (\ref{expan}).

\begin{figure}[b]
\centerline{
\includegraphics[width=9.5cm]{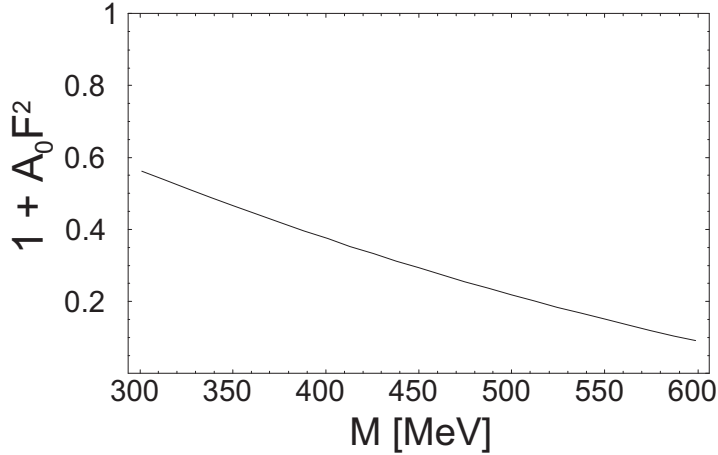}} 
\caption{The quantity $1+A_0 F^2$ evaluated in the Nambu--Jona-Lasinio
model with the sharp four-momentum regulator, plotted as a function 
of the quark mass, $M$.
Large deviation from unity is found. The calculation is performed in the 
chiral limit, and with the cut-off chosen in such a way as to 
reproduce the physical value of the pion decay constant. }
\label{fig1}
\end{figure}

Note that while the sign of $Z_{0}$ is positive-definite, the sign $A_{0}$
is negative-definite. Therefore the presence of the $A$-term in the effective
Lagrangian lowers the wave function renormalization of the $\sigma$-field,
thereby increasing its mass, as follows from Eq. (\ref{ms}).

\section{Chiral solitons}\label{sec:sol}

In the following parts of this paper we are going to explore the dynamical
consequences of the $A$-term for chiral solitons. First, we consider the
simplest case: the 
generalization of the Gell-Mann--L\'evy Lagrangian of the form 
\begin{eqnarray}
L &=&\frac{1}{2}[(\partial _{\mu }\sigma )^{2}+(\partial _{\mu }\pi ^{a})^{2}%
]+\frac{1}{2}A_{0}[\sigma \partial ^{\mu }\sigma +\pi ^{a}\partial ^{\mu
}\pi ^{a}]^{2}-\frac{\lambda ^{2}}{4}(\sigma ^{2}+\pi ^{2}-\nu
^{2})^{2}\nonumber \\ &+& c\sigma   
-g\bar{\psi}(i \slashchar{\partial} +\sigma +
i\gamma _{5}\tau ^{a}\pi ^{a})\psi ,
\label{sim}
\end{eqnarray}
with 
\begin{eqnarray}
\lambda ^{2} &=&\frac{\bar{m}_{\sigma }^{2}-m_{\pi }^{2}}{2F^{2}},\quad \nu
^{2}=\frac{\bar{m}_{\sigma }^{2}-3m_{\pi }^{2}}{\bar{m}_{\sigma }^{2}-m_{\pi
}^{2}}F^{2},  \quad
c =m_{\pi }^{2}F, \nonumber \\ \bar{m}_{\sigma }^{2}
&=&(1+F^{2}A_{0})m_{\sigma }^{2}.
\end{eqnarray}
The last equality ensures that the sigma mass is $m_{\sigma }$, according to
Eq. (\ref{ms}). The constants $g$, $m_{\sigma }$, and $A_{0}$ are treated as
model parameters. The resulting mean-field equations of the form (\ref{ELmat}%
), and the equations for the upper and lower components of the valence quark
orbital \cite{BirBan8485,KRS84}, are solved with the hedgehog ansatz 
for the chiral
fields: $\sigma (x)=\sigma (r)$, $\pi ^{a}(x)=r^{a}/r$ $\pi (r)$, where $r$
denotes the radial coordinate.

\begin{figure}[b]
\centerline{
\includegraphics[width=11.5cm]{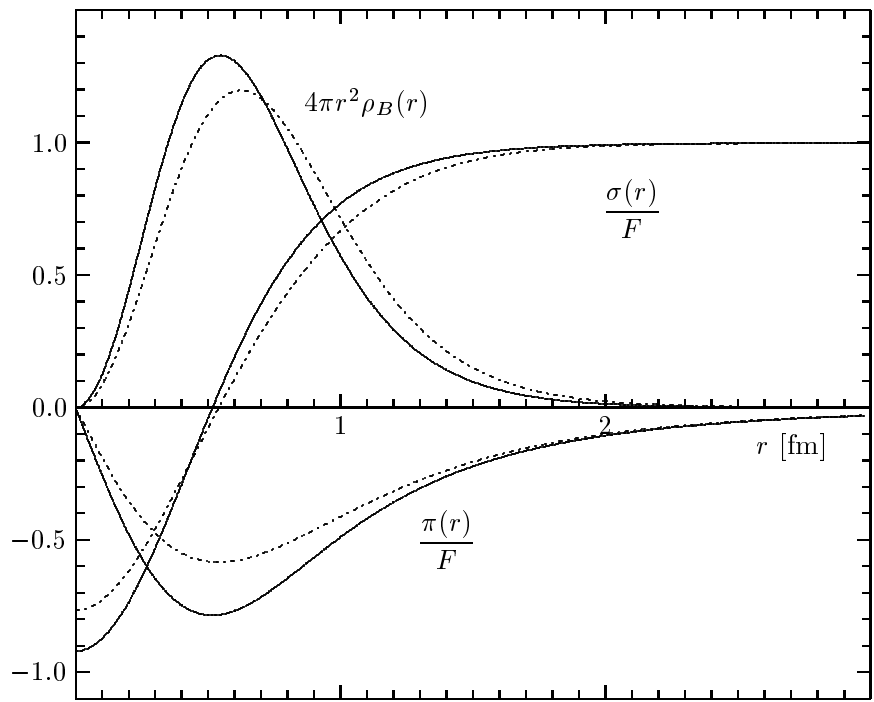}}
\caption{The chiral fields and the radial quark density in 
the hedgehog chiral soliton
of model (\ref{sim}), 
plotted as functions of the radial coordinate  $r$. 
Two solutions are compared: with  $A_0=0$ 
(solid lines) and $A_0 F^2=-0.6$ (dashed lines). 
The remaining model parameters are  $g=4.5$ 
and $m_\sigma=900$~MeV.}
\label{fig2}
\end{figure}

A typical solution, for $g=4.5$, $m_{\sigma }=900$~MeV, and $F^{2}A_{0}=-0.6$
(dashed lines) and $A_{0}=0$ (solid lines),  is presented in Fig. 2, where
we show the chiral fields, and the radial valence quark density (radial
baryon density), $4\pi r^{2}\rho _{B}(r)=4\pi r^{2}q^{\dagger}q$, plotted as
functions of the radial coordinate $r$. The chosen value of $A_{0}$ 
is suggested by the analysis
of Sec. \ref{sec:motiv}, \emph{cf.} Fig. 1. 
The values of $g$ and $m_{\sigma }$ are the
typical values used in other studies. We notice that the presence of the $A$%
-term weakens the pion field by about $20\%$, modifies the shape of the $%
\sigma $ field, as well as slightly increases the size of the soliton, which
is visible from the baryon-density curves. The results do not depend
qualitatively on the values of the parameters $g$ and $m_{\sigma }$. 
As expected, the
effect is more pronounced
if the chiral fields lie away from the chiral circle.

\begin{figure}[b]
\centerline{
\includegraphics[width=11.5cm]{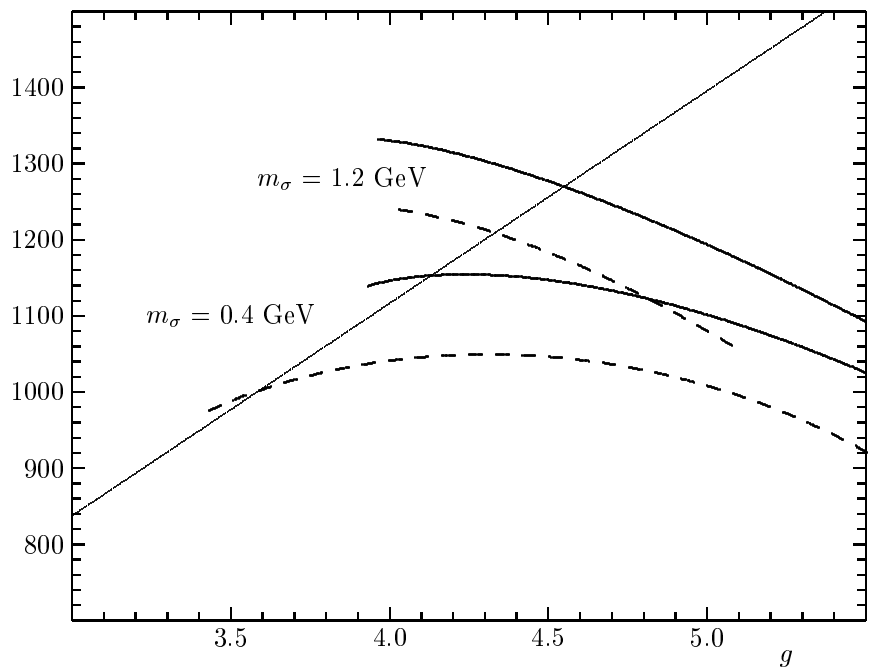}}
\caption{The dependence of the energy of the soliton on 
the coupling constant $g$ for 
$A_0=0$ (solid lines) and $A_0 F^2=-0.6$ (dashed lines). 
Two values of $m_\sigma$ are used.
The straight solid line is the energy of three free quarks. 
The presence of the $A$-term 
noticeably lowers the energy of the soliton.}
\label{fig3}
\end{figure}

In Fig. 3 we study the behavior of the soliton energy as a function
of the coupling constant $g$. Again, we use $F^{2}A_{0}=-0.6$ (dashed lines)
and $A_{0}=0$ (solid lines). For $m_{\sigma }$ we take two ``extreme
values'', $400$~MeV and $1200$~MeV.  We note that the inclusion of the $A$%
-term lowers the energy of the soliton for all values of $g$. The lowering
is a significant effect, by the amount of a few tens of MeV. The
solitons become energetically stable when their energy is lower than the
energy of three free quarks, $3M$, denoted in Fig. 3 by 
the straight solid line. 
With the $A$-term present, the stability is achieved at lower values of $g$. 
This is a desired effect since many phenomenological approaches
\cite{GolliRosina,Lubeck,Birse85,NeuberGoeke,Amoreira,drago}
have problems in getting the mass in the right ball park
if $g$ is too high.

On the other hand, as expected, the $A$-term has negligible effect on the
nucleon observables (only of the order of few percent), 
since it does not explicitly 
enter in the expressions for the Noether currents (\ref{VA}).

\section{Model with the non-local regulator}\label{sec:nonloc}

Recently, considerable progress has been made in the treatment of chiral
quark models with non-local interactions. It these models, rather then
cutting the momentum in the quark loop, the finiteness of the theory is
achieved by the non-locality of the four-quark interaction. 
In fact, the model derivations of effective chiral quark models, such as the 
instanton-liquid model \cite{mitja:rev,instant1:rev}, 
or the resummations of rainbow diagrams
in the Schwinger-Dyson approach \cite{Roberts:rev,RobWil:rev}, 
lead in a natural way to 
non-local models \cite{ripka:book}. Their applications range from the 
meson sector \cite{PlantB,BowlerB,coim99wb,Plant:ml}, 
through quark matter at non-zero temperature
\cite{basz} and baryon number \cite{Carter,scoc2,scoc1,gocke},
chiral solitons \cite{solnonl,solilong}, to applications in exclusive 
processes in QCD \cite{Michal1,Dorokhov:nl}.
The advantages of non-local models include such features as 
the natural preservation of anomalies, finiteness to all orders in the
$1/N_c$-expansion, such that the meson-loop calculations can be 
done with no extra parameters \cite{Plant:ml,basz,Ripka:ml}, 
or the stability of solitons without 
the need for the non-linear constraint.
The solitons of non-local models, their construction and the resulting 
phenomenology, have been discussed in detail in Refs. \cite{solnonl,solilong}.
Here we are going to check how well we can approximate the full, 
complicated model, with
the $\sigma$-type model supplied with the $A$-term. 

The effective action for the non-local model has the form (\ref{Action}), 
with the crucial
difference that now the Dirac Hamiltonian carries the regularization 
operator, $r$ 
\cite{ripka:book}: 
\begin{equation}
h =-i\vec{\alpha}\cdot \vec{\nabla}+r(\partial^2) \beta g(\sigma 
+i\gamma _{5}\tau
^{a}\pi ^{a}) r(\partial^2)+\beta m.  \label{hamnl}
\end{equation}
The current quark mass is denoted by $m$.
It is convenient to introduce the notation $R=r^2$.
One can straightforwardly derive the expression for the effective potential,
\begin{eqnarray}
V(\sigma ,\pi ^{a}) &=&-2N_{c}N_{f}\int \frac{d_{4}k}{(2\pi )^{4}}
\times \\ && \left[
\log \left( 1+\frac{g^{2}R^{2}\delta +2mgR(\sigma -F)}{k^{2}+\left(
RM+m\right) ^{2}}\right) -\frac{g^{2}\left( R^{2}+Rm/M\right) \delta }{%
k^{2}+\left( RM+m\right) ^{2}}\right]   \nonumber  \label{v2mfin} \\
&=&V_{2}(\sigma ^{2}+\pi ^{2}-F^{2})^{2}+\frac{1}{2}m_{\pi }^{2}\left[
(\sigma -F)^{2}+\pi ^{2}\right] +O(m^{2},m\delta ^{2},\delta ^{3}), 
\nonumber 
\end{eqnarray}
which gives  
\begin{equation}
V_{2}=N_{c}N_{f}g^{4}\int \frac{d_{4}k}{(2\pi )^{4}}\frac{R^{4}}{\left(
k^{2}+R^{2}M^{2}\right) ^{2}}.  \nonumber
\end{equation}
The formulas for $Z_0$ and $A_{0}$ obtained from the non-local model read
\begin{eqnarray}
Z_0 =2N_c N_f g^2 \int \frac{d_4k}{(2\pi )^4}\frac{R^2-k^2RR^{\prime
}+(k^2R^{\prime })^2}{\left( k^2+R^2 M^2 \right) ^2}, 
\label{Z0fin}
\end{eqnarray}
\begin{eqnarray}
A_{0}&=&-8N_{c}N_{f}g^{4}\int \frac{d_{4}k}{(2\pi )^{4}} \times \nonumber \\
&&\frac{R^{2}\left(
(k^{2}+2R^{2}M^{2})R^{2}-2k^{2}(2k^{2}+R^{2}M^{2})RR^{\prime
}+8k^{6}R^{\prime }{}^{2}\right) }{4\left( k^{2}+R^{2}M^{2}\right) ^{4}},
\label{A0fin}
\end{eqnarray}
where $R'=dR/d(k^2)$. 
The model parameters are fitted in such a way that $Z_0=1$, which ensures that 
$F$ has its physical value \cite{PlantB,BowlerB,solilong}.

\begin{figure}[b]
\centerline{
\includegraphics[width=11.5cm]{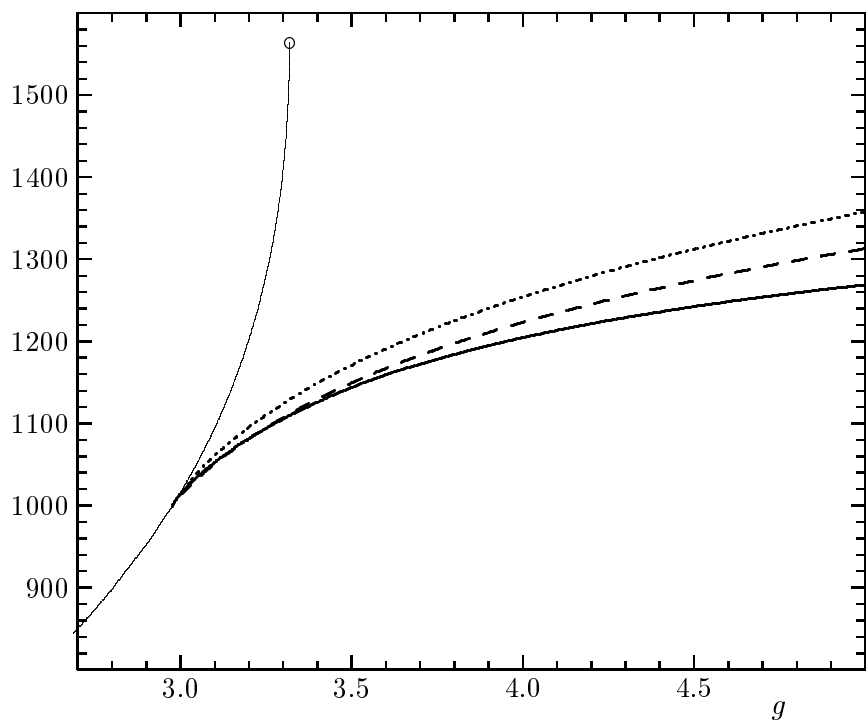}}
\caption{The dependence of the energy of the soliton of the model with
non-local regulators on the coupling constant $g$. The solid line on the right
corresponds to the solution of the 
quark model with non-local regulators, the dotted line shows the solution 
of the approximating $\sigma$-like model with $A_0=0$, and the dashed line 
shows the solution of the approximating model with the $A$-term present. 
The solid line on the left corresponds to the energy of three free quarks
in the non-local model.}
\label{fig4}
\end{figure}

The presence of the regulator $r$ dependent on the momentum modifies 
the Noether currents
of the theory \cite{solnonl,solilong,bled99}. As a result, discussed in 
detail in Refs. 
\cite{solnonl,solilong}, the contributions of the valence quarks to 
the Euler-Lagrange 
equations carry an extra residue factor, 
$z_{\rm val}$. This factor is defined as 
\begin{equation}
z_{\rm val}=\left ( 1-i \left .  
\frac{d \epsilon_{\rm val}(\omega)}{d\omega} \right | _{\omega=i 
\epsilon_{\rm val}}    
\right )^{-1},
\label{zval}
\end{equation}
with $\epsilon_{\rm val}(\omega)$ denoting the energy-dependent 
valence-orbital eigenvalue
of the energy-dependent Dirac Hamiltonian (\ref{hamnl}). It 
satisfies the equation 
\begin{equation} h\left( -\epsilon_{{\rm val}}^{2}\right) |{\rm %
val}\rangle =\epsilon_{{\rm val}}|{\rm val}\rangle .
\label{eval}
\end{equation} 
Thus, 
\begin{eqnarray}
j_{\sigma }(\vec{x}) 
&=&
gN_{c}z_{\mathrm{val}}\langle \mathrm{val}|r|\vec{x}
\rangle \beta \langle \vec{x}|r|\mathrm{val}\rangle , 
\\
j_{\pi}^{a}(\vec{x}) 
&=&
gN_{c}z_{\mathrm{val}}\langle \mathrm{val}|r|\vec{x}
\rangle \beta i\gamma _{5}\tau ^{a}\langle \vec{x}|r|\mathrm{val}\rangle . 
\label{jz}
\nonumber
\end{eqnarray}

In the numerical study presented in this paper we use the Gaussian regulator,
\begin{equation}
R(k)=r(k)^2=\exp (-\frac{k^{2}}{2\Lambda ^{2}}). \label{regu}
\end{equation}

In Fig. 4 we compare the hedgehog soliton solutions of the 
non-local quark model, 
defined by Eqs. (\ref{Action},\ref{hamnl},\ref{regu}), 
with those of the approximating 
$\sigma$-like model, defined by 
Eqs. (\ref{Lag},\ref{v2mfin},\ref{A0fin},\ref{jz}),
and prescriptions ({\ref{zval},\ref{eval},\ref{jz})
for the valence orbit. We plot
the energy of the solution as a function of the quark-meson 
coupling constant, $g$.
The solid line on the right corresponds to the non-local quark model, 
the dotted line to the 
approximating model with $A_0=0$, and the dashed line to 
the approximating model with 
$A_0$ given by Eq. (\ref{A0fin}).
Similarly to the study of Sec. \ref{sec:sol}, the inclusion of 
the $A$-term results in 
lowering of the energy. We note that the model with the 
$A$-term approximates the full
non-local quark model better than the model with $A_0=0$. 
In particular, at lower values 
of $g$ the agreement is very good, since in that region the solitons are 
large and the gradient expansion works well even at the lowest order. 
The solid line on the left corresponds to the energy 
of three free quarks in the non-local quark model \cite{solnonl,solilong}.

Comparing the nucleon observables calculated in the approximating model
to those calculated in the full model \cite{solilong} we find that
the valence contribution to $g_A$ and the magnetic moment is
reproduced within a few percents for $g$ below 3.5 (corresponding to 
$M$ below 325~MeV), and within 10~\%
for $g\ge 3.5$. 
The meson contribution to the magnetic moment is again reproduced well
for $g\sim 3.3$, and is typically 20~\% larger
than the sea contribution in the quark model for $g\sim 3.7$,
and more than 30~\% for $g\ge 5$.
On the other hand, the meson contribution to $g_A$ is a factor of 2
too large compared to the sea contribution already for $g$ slightly 
above the threshold value $g=3$;
this factor rises to 3 at $g\sim 5$.
The results do not change much if we switch-off the $A$-term;
the only noticeable difference is in the meson contribution
to the magnetic moment at larger values of $g$ where
the inclusion of the $A$-term improves the results by
5~\% -- 10~\%. This is a dynamical effect reflecting the tendency of
the $A$-term to reduce the strength of the pion field, as can be seen already
in Fig.~\ref{fig2}.

\section{Conclusion}\label{sec:con}

We have analyzed the models approximating the chiral quark models with 
four-quark interaction. The models are constructed with help of 
the gradient-expansion
technique, good for large solitons. We have found that the inclusion of the 
$A$-term, appearing naturally in the gradient expansion of the
underlying theory, improves the 
approximating model. The term results in a significant (a few tens of MeV) 
lowering of the energy of the soliton. 
This effect is particularly important
in the non-local model, where the $A$-term considerably
improves the approximating model such that for lower values of
the quark-meson coupling
the energy follows almost exactly the energy obtained
in the full quark model.
We wish to stress that the variants of the $\sigma$-model, which result 
from the approximation of the full quark model via the gradient expansion are 
much simpler to solve than the original theory. 
They avoid the complications of the numerical 
treatment of the Dirac sea, and thus are useful and more practical in
the studies of soliton models of baryons.    

\bigskip

The authors wish to thank Enrique Ruiz Arriola and George Ripka for 
several useful discussions. One of us (WB) wishes to thank 
the organizers of the 
Bled Workshops in Physics, where this research was initiated, for their
kind hospitality and stimulating atmosphere.

\end{document}